# Application of Sizing Estimation Techniques for Business Critical Software Project Management

[*1] Parvez Mahmood Khan, [2] M.M. Sufyan Beg
Department of Computer Engineering, J.M.I. New Delhi, India
[1] pmkhan@hotmail.com , [2] mmsbeg@cs.berkeley.edu

*Abstract* Estimation is one of the most critical areas in software project management life cycle, which is still evolving and less matured as compared to many other industries like construction, manufacturing etc. Originally the word estimation, in the context of software projects use to refer to cost and duration estimates only with software-size almost always assumed to be a fixed input. Continued legacy of bad estimates has compelled researchers, practitioners and business organizations to draw their attention towards another dimension of the problem and seriously validate an additional component – size estimation. Recent studies have shown that size is the principal determinant of cost, and therefore an accurate size estimate is crucial to good cost estimation[10]. Improving the accuracy of size estimates is, therefore, instrumental in improving the accuracy of cost and schedule estimates. Moreover, software size and cost estimates have the highest utility at the time of project inception - when most important decisions (e.g. budget allocation, personnel allocation, etc). are taken. The dilemma, however, is that only high-level requirements for a project are available at this stage. Leveraging this high-level information to produce an accurate estimate of software size is an extremely challenging and high risk task. This study acknowledges the presence and effect of risk in any software estimate and offers pragmatic strategies for risk mitigation.

* Corresponding Author:
Parvez Mahmood Khan,
Department of Computer Engineering, J.M.I.
New Delhi, India
Email: pmkhan@hotmail.com

## 1. Introduction

Since the beginning of computer era, software project estimations continue to be an important and challenging task in software development projects. In the early days of computers, hardware costs were really high and software costs were relatively lower. Consequently, impacts of bad estimates for software development were not so pronounced. With reduction in hardware prices, advent of Personal Computers and growth in software costs, the impact of bad software estimates started drawing attention of organizations, practitioners and researches more than ever before. Challenges associated with the task of estimations, have multiplied many folds in recent past, on account of well know reasons like:

    (i)    Increase in Size of Applications

  (ii)  Increase in Complexity of Applications
  (iii)  Diversification of Technologies (plenty of technology options)
  (iv)  Customer demand for predictability of software project cost (Accuracy of Cost Estimates)
  (v)  Customer demand for predictability of project delivery schedule (Accuracy of Schedule Estimates)
  (vi)  Performing organization's internal mandates to improve efficiency of project team (Need to Control and Reduce Effort Overruns)

Recent global recession and economic slowdown in US and Europe markets has only *added insult to the injury* by worsening the problems caused by poor estimates, and created additional pressures on Indian IT-companies, because most of the customers and revenues for many of Indian IT-companies were historically linked to US and Europe markets[11]. On an average, 35% of the historical(pre-recession period) revenues for Indian IT-Companies
were linked to overseas customers in BFSI-sector, which was worst hit BUs/Verticals within the IT-Companies were BFSI-Verticals (due to collapse of many banks in US-market). Other BUs were also impacted, forcing many Indian

IT-Companies to start looking for additional business opportunities in other Geographies (like Middle East, APAC, within India etc.). Apart from the business development initiatives (like exploring other geographies, giving additional customer incentives etc.), organizations are also willing to commit to delivering Fixed-Price Projects, despite understanding higher risks associated with fixed-price projects. Distressingly, recent industry trends have found that majority of the projects exceed budget, are completed past scheduled deadlines and do not meet original business objectives[12].

This paper is the result of an study undertaken with the research objective of examining the *role of the application of sizing estimation techniques* in the wide-spread poor performance of business critical software projects (particularly fixed-price projects) in the real world software projects and propose possible solutions to minimize impact of software crisis. It is found that one of the significant factors contributing to the chaos is poor estimates for software projects.

## 2. Estimation Problems Of Business Critical Software Projects

An estimate is an educated prediction of how long a project will take and how much will it cost. It is a tentative evaluation or rough computation. Most of the time, it is subjective and based on an individual's (practitioner's, researcher's, project team's) experience, perception and opinion. Generally speaking, an estimation model can be of two types: direct estimation or derived estimation. In a direct estimate, the practitioner tries to go directly to the final results that he wants after examining all the available information and documentation ( like product descriptions, requirements, process details, assumptions, constraints etc.). Very experienced practitioners can be very good at making direct estimations, but they are often unable to document and explain how did they arrived at the results of the estimation exercise. Consequently, it is not easy to share the results with other people or to make a business case based on the estimate. In contrast, a derived estimate employs a method that is a step-by-step written or modeled process. It starts with input points and, by means of an algorithmic model, produces a final estimate. There are many standard as well as proprietary models are available in the industry. These models are based on historical data, which can be either local historical data or general world market databases such as the ISBSG.

### 2.1 Fundamental Dilemma of Early Estimates in Business Critical Software Projects:

The beginning of all business critical software development projects is the point in time when performing organizations are least certain about the project, yet it is also the time when top management expects to be delivered project estimates that are very precise. The difficulty of accurate size estimation is compounded especially at the time of inception when very little information is available. This problem is visually depicted in the "cone of uncertainty" diagram which was initially conceived by Boehm and subsequently published by S. McConnell [13], which is reproduced in Figure 1. This diagram plots the relative size range versus the various phases and milestones. It clearly indicates that, at the beginning of a project, estimates are likely to be far off from the actual values. As time progresses, the error in estimation starts decreasing but so does the utility of these estimates.

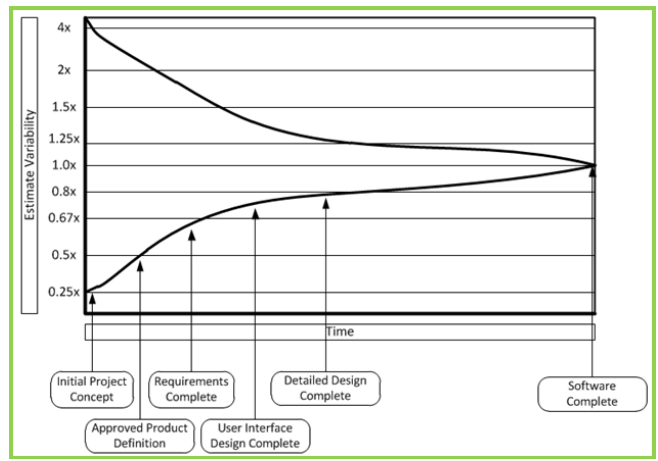

**Figure-1:** *Cone of Uncertainty in Software Project Estimation* – adopted from [13]

The cone of uncertainty highlights the fundamental dilemma of system and software size estimation i.e. the tradeoff between accuracy and utility of estimates. Estimates are of highest value at the beginning of a project. This is the time when important decisions such as those regarding budget and personnel allocation are made. Estimation accuracy at this point, however, is lowest. Therefore, in order to achieve maximum utility, it is extremely important to improve the accuracy of estimates produced at the time of inception.

As evident from figure-1, maximum uncertainty is at the beginning of any project (a variance of 0.25x to 4x in range). This variance means that, if we estimate a project schedule to be one-year, it could actually end-up taking anywhere between 3-months to 48-months. Similarly, if we estimate a project cost to be 1Million USD (at inception of the project), there is a possibility that we could actually end-up spending anywhere between 25,000 USD to 4 Million USD by the end of the project.
All software projects are subject to inherent errors in early estimates. The Cone of Uncertainty[13] represents the best-case reduction in estimation error and improvement in predictability over the course of a project. We need to treat the cone as a fact of life and plan to manage business critical software projects accordingly.

   **2.2 Issues with Existing Sizing Measures for Estimation of Business Critical Software Projects:**
Strictly speaking, physical size of an object is something that should not change. For example:

Size of the room we are sitting in does not change irrespective of the person measuring the size is more experienced or less experienced.

Similarly, weight of 1-kg of vegetables remains constant even if the measuring scale is mechanical scale or an electronic scale.

The distance of one kilometer remains one kilometer, irrespective of whether a kid is walking or an elderly person is walking.

Converting data from Kilograms to grams (or vice versa), results are same irrespective of the experience, age, language and tool used for conversion of data across multiple units of measures.

Unfortunately, in software engineering, we are still having a long-long way to go before we can reach that level of maturity in measurements[15],[16].

Given the understanding that a software sizing measure is fundamental to any software measurement program, various sizing measures, can be broadly classified into two-categories:

Size of Source Code (SLOC) Measurements

Functional Size Measurements: These can be sub-categorized as:

Function Points & Feature Points
IFPUG FPA
NESMA FPA
COSMIC-FFP
FPA Mark-II
Object Points
Use Case Points
Internet Points
Domino Points
SAP Points
etc.

Despite on-going usage of both SLOC(Source Lines of Code) as well as FP(Function Points) as valid software sizing measures in real world software developments in IT-industry, the two actually measure different things and have very different characteristics:

SLOC is a measure of the size of the system that is built. It is highly dependent on the programming environment, language and technological aspect used to build the system. There are many well-documented problems and issues with LOC. There is no agreement on how to count Lines of Code – logical statements or physical statements, treatment of inline documentation., etc. SLOC estimates are prone to errors especially for early size estimations when technology details are mere assumptions and educated guess. Despite these problems, LOC is still frequently used by very reputable and professional organizations in IT-industry.

In contrast to SLOC, function points (FP) is a measure of required functionality, and is relatively independent of the technology used to develop the system. While FP addresses many of the problems inherent in SLOC, specially for early size estimates and is witnessing unprecedented growth in it's adoption and usage, still it has its own set of issues and improvement opportunities [6],[7], [14].

Despite theoretical claims on availability of backfiring techniques, there is limited evidence of actual deployment of any standard way of converting software size from one measure to another, across the business organizations. There are issues and practical challenges in converting software sizing estimates data from SLOC to FP or vice versa.

Besides this, another aspect of these measures is that the size is adjusted (increase or decrease) due to factors of complexity etc. which adds another element of subjectivity, making final size estimates error prone.

In this research work, an attempt is made to study and analyze possible impacts of application of sizing estimation techniques for business critical software projects and empirically assess if there is any correlation of the same with project performance/outcomes.

### 3. Research Background And Context

Published results from industry oriented research encompassing real-world software projects data, continues to support the notion that a "software crisis" continues unabated. Software projects continue to run behind schedule and over budget, as has been reported for years [12].

**Table-1:** Business Critical Software Project Outcomes in Global IT-Industry [12]

| Software-Projects | 1994 | 1996 | 1998 | 2000 | 2002 | 2004 | 2009 |
|---|---|---|---|---|---|---|---|
| Succeeded | 16% | 27% | 26% | 28% | 34% | 29% | 32% |
| Challenged | 53% | 33% | 46% | 49% | 51% | 53% | 44% |
| Failed | 31% | 40% | 28% | 23% | 15% | 18% | 24% |

Companies continue to invest a higher and higher percentage of revenues into IT development as they incorporate automation into more of their operations. Consequently, an increasing amount of money is being put into IT development, but companies are still routinely not achieving good value. Systems cannot be counted on to finish on time or within budget, implement all of the required features or deliver the necessary quality requirements, despite
availability of many estimation models & techniques and software engineering processes. Even the best estimates for business critical software projects compiled on large scale projects continue to fall short of actual results.

From figure-2 it is clearly evident that, a vast majority of the projects are either failed or challenged with poor performance. In this research work, an attempt is made to study and analyze possible impacts of application of sizing estimation techniques for a good sample of failed and challenged projects from It-Industry and empirically assess if there is any correlation of the sizing estimation technique used on those projects with project performance/outcomes. Result of this research is envisaged to benefit both IT-companies which are having their core business of undertaking software development projects for their customers as well as in-house application development teams within large business organizations.

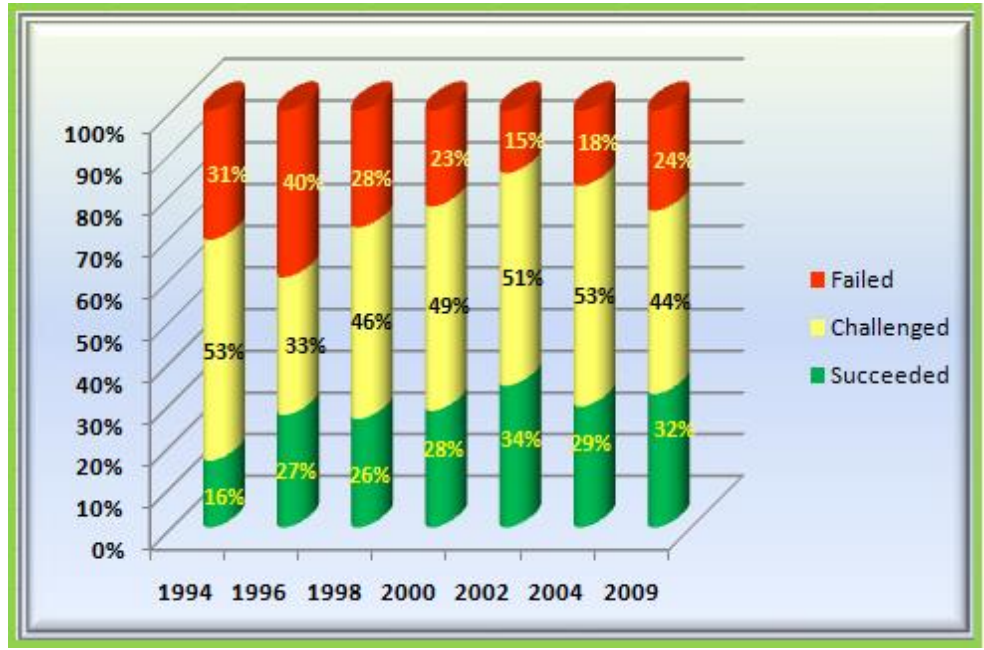

**Figure-2:** *Trend Analysis of Business Critical Software Project Outcomes in Global IT-Industry*

## 5. Observations On Related Work

The attempt to gain a more complete understanding of nuances of software project estimations for business critical software projects has been of great interest to researchers, practitioners as well as to all professionally managed business organizations engaged with Software Projects. Thorough understanding of estimation process components, thereby improving the existing estimation practices can help business organizations in overall project delivery capability and maturity within the context of organization – thereby facilitating achievement of strategic goals & objectives, for which the business critical projects were undertaken. Several lines of research exist in the growing body of literature dealing with the subject, all of it an attempt to get a solid understanding of various attributes of estimation process[1],[5], [8].

In 1979, Allan J. Albrecht of IBM published a new conceptual groundwork by proposing the idea of sizing the software applications purely from the point of view of functionality[2] and envisaged to be technology agnostic. This new method described by Allan J. Albert was named Function Point Analysis(FPA) as it conceptualized the idea of quantifying the size by measuring amount of functionality in terms of Function-Points(FPs). The initially proposed method in 1979 was further improvised by Albrecht along-with Gaffney in 1983[3] and published again in

November 1984[4]. Many practitioners and researchers, have since then made significant attempts to measure functional size, but Albrecht's work is still considered as the key foundation which opened many lines of research in the field of software size estimations using the functional size measurement philosophy.
Consequent upon the conceptual groundwork of Albrecht from IBM in 1979[2], several functional size measurement methods were proposed, and are still evolving. An overview of the same is depicted in figure-3.

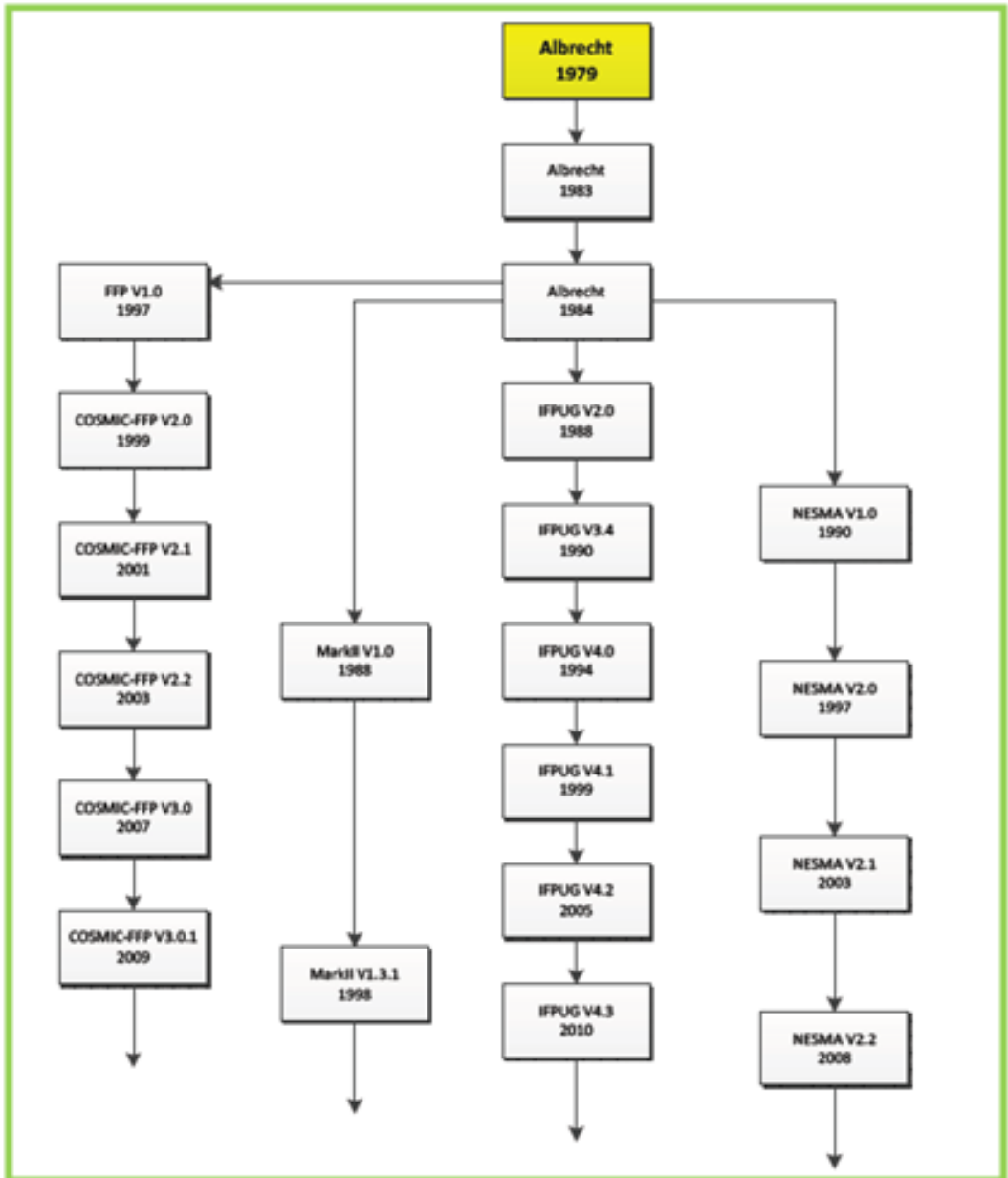

**Figure-3:** *Overview of FSM Methods Evolution since conceptual groundwork of Albrecht of IBM, in 1979*

In 1986, a professional, non-profit organization called International Function Point User Group (IFPUG)[9] was formed to promote, maintain and advance the practice of using Function Point metric in the Global IT-Industry in-line with Albert's original vision of utility of function point metrics.

IFPUG has modified Albrecht's original method several times and published its own Counting Practices Manual's (CPM) to clarify and standardize rules for the application of Functional Point Analysis techniques. Chronological order of Counting Practice Manuals released by IFPUG is listed in Table-2.

**Table-2**: Major Releases of IFPUG's Counting Practices Manuals, so far

| Sr. No. | IFPUG's Publications | Formal Release Timeline | Remarks |
|---|---|---|---|
| 1 | IFPUG's CPM – Release 2.0 | 1988 | (i) In these CPMs, Function Point Analysis is formulated as a counting method of several steps. (ii) Each release of these IFPUG publications contained progressive refinements to the technique originally presented by Albrecht, but there is no shift from the initial conceptual framework and it is still very close to Albercht's original publication despite more than 3-decades of progressive elaboration in counting practices. |
| 2 | IFPUG's CPM – Release 3.0 | 1990 | |
| 3 | IFPUG's CPM – Release 4.0 | 1994 | |
| 4 | IFPUG's CPM – Release 4.1 | 1999 | |
| 5 | IFPUG's CPM – Release 4.2 | 2005 | |
| 6 | IFPUG's CPM – Release 4.3 | 2010 | |

The IFPUG's CPM is a body of knowledge used by function point analysts to measure the functional size of applications and projects for benchmarking and estimating world-wide for many domains and business areas. The CPM is an internationally approved standard under ISO/IEC 14143-1 Information Technology – Software Measurement. Latest revision of Counting Practices Manual (i.e. CPM 4.3) which was adopted w.e.f. 1$^{st}$ January 2010 (replacing CPM 4.2) further clarifies the rules and enhances the definitions and examples, thereby enabling a more consistent interpretation and application of rules.

In another research[22], effort is made to quantify IT Estimation Risks, so that necessary mitigation actions can be planned for different types of estimation methods. There exist different types of Functional Size Measurement Methods and others are still evolving. All the existing FSM methods measure the functional size by means of the "functionality" delivered to the users, the main differences between these techniques arise from what they count and how they do it. As of now, four methods are certified by ISO as an international standard[18]-[21], viz. IFPUG FPA, Mk II FPA, COSMIC FFP and NESMA FSM.

In related research works[23], focus is on kicking-off risk management for software projects, which is relevant to our study, in terms of need to mitigate risks associated with software sizing estimations. In [24] value propositions of PMO are discussed and [25] suggests measurements of cost components related to quality for effective SQA. In [26], need for increasing the objectivity of SDLC-model selection is discussed and a decision support matrix is proposed.

As noted above, little work has been devoted to objectively evaluate the impact of wrong or insufficient application of sizing estimation techniques on the outcome of business critical software projects. This research paper is an attempt to address this issue, by making an assessment of possible impacts of insufficient or incorrect application of sizing estimation techniques on business critical software projects any itself and proposing a layered approach to application of sizing estimations to business critical software projects with a view to alleviate the software crisis.

## 6. Research Methodology

Grounded theory research methodology was chosen for this study with a view of qualitative analysing of application of sizing estimation techniques for business critical software project management. There were two reasons for adopting such a method for this research:
  (i) First, the research was aimed at the extension of existing theory. Grounded theory is generally deemed appropriate for such efforts as it allows theory to emerge from interview transcripts and organization artifacts.
  (ii) Second, the methodology is reputed to help separate researcher biases from interpretation of the data.

Study was done in two-phases. During Phase-I, a survey of real world project managers and practitioners from IT-Industry covering multiple organizations and different geographies conducted in parallel, over multiple professional meets and technical forums to acquire real life experience of senior project managers on their real world experiences. Questions posed to the participants of the survey is listed below in Table-3. A total of 59-project managers and practitioners participated in the survey and their objective responses are tabulated in Table-4.

**Table-3:** Questions Used in Survey of Practitioner's

| | |
|---|---|
| **In your experience with real world projects, how many times – do you believe that project delivery teams (engaged in software development projects) actually perform detailed sizing estimations:** | |
| **Option-1** | Only once (at the time of proposal/bidding for the project) |
| **Option-2** | Twice (First at the time of proposal/ bidding and then at the time of closure of the project) |
| **Option-3** | At least Thrice (At inception, closure and whenever there is a scope change request) |
| **Option-4** | Never Required (Only Cost-Estimates and Schedule-Estimates are worked-out) |
| **Option-5** | Depends on customer's requirement for detailed size-estimates to be submitted |

**International Journal of**
**Soft Computing And Software Engineering (JSCSE)**
e-ISSN: 2251-7545
Vol.3,No.6, 2013                                                    DOI: **10.7321/jscse.v3.n6.2**
Published online: Jun 25, 2013
During Phase-II, a detailed study and thorough assessment of 11- real-life, global software development /SW-implementation projects, was undertaken with the objective of carrying out qualitative analysis of sizing estimation for business critical software projects.

Project documents of all the projects selected for this detailed study - including Project Charted, Scope Document/SRS, Detailed Estimation Worksheets, Non-transactional Project-Plan, Transactional Project Plan, PDMRs, monthly PMRs, monthly presentations made to PMOs (wherever applicable) and monthly reports formally communicated to project sponsor and key stakeholder(s) were thoroughly examined. In the event of any ambiguities and lack of organizational processes, standard best practices of PMBOK[17] were used as the baseline for gap analysis. Necessary clarifications were also sought from project manager/project team/PMO during the course of project document reviews. Authenticity of reports were reconciled with PMIS-data and a few inconsistencies observed (during reconciliation process) were challenged with reference to the project life cycle documentation - as and when required during the course of study. One of the key objectives of study was to understand the process followed by the projects to arrive at *cost-estimates* and *schedule-estimates* and utility of *sizing estimates* in each of the projects under study. An special emphasis was given on analysing the approach of project teams on managing risks related to estimation techniques, process and life cycle stage of the project when the estimations were performed.

A total of 37 persons, including project sponsors, program managers, project managers, business analysts, senior team members of project team and PMO-staff (related to the projects selected for detailed study) were interviewed over a period of 13-months. The persons chosen for the interview were generally very experienced in their position with average experience of more than 12-years and had previous experience of managing project budgets of multi-million-dollors. One of the project managers interviewed was also ex-employee, having left the organization within the preceding 3-months of the study.

The interviews were facilitated via open-ended questions intended to figure-out project issues arising out of *underestimations* and *overestimations* and also capture ideas, suggestion and feedback from project practitioners, business analysts and project sponsors. The resulting pages of interview transcripts were analysed to identify recurrent themes and concepts associated with underestimations and overestimations – across the projects. Those themes and concepts were then tested against archived data including survey results, with an eye to find evidence reinforcing or challenging the themes and concepts. Preliminary conclusions were then tested via follow-up interviews with a subset of original 32-individuals. This iterative process (interviews, concepts identification, concepts aggregation, theme analysis, testing against organizational artifacts, review with interviewees, adaption, repeat) led to the identification of key concepts and vital lessons learnt for estimations related to business critical software projects. Most important learning experiences from this study are articulated in the following sections 7 and 8.

### 7. Discussion On Our Research Findings

Summary of responses captured from survey of practitioners and project team members is tabulated in Table-4 and plot of survey results is depicted in Figure-4.

28

Table-4: Summary of responses captured in Survey of Practitioner's

| Options articulated | No. of Responses | % of Responses |
|---|---|---|
| Only once (*at the time of proposal/bidding for the project*) | 31 | 53 % |
| Twice (*First at the time of proposal/ bidding and then at the time of closure of the project*) | 20 | 34 % |
| At least Thrice (*At inception, closure and whenever there is a scope change request*) | 05 | 08 % |
| Never Required (*Only Cost-Estimates and Schedule-Estimates are worked-out*) | 01 | 02 % |
| Depends on customer's requirement for detailed size-estimates to be submitted | 02 | 03 % |

**Total responses → 59**
---------------------------

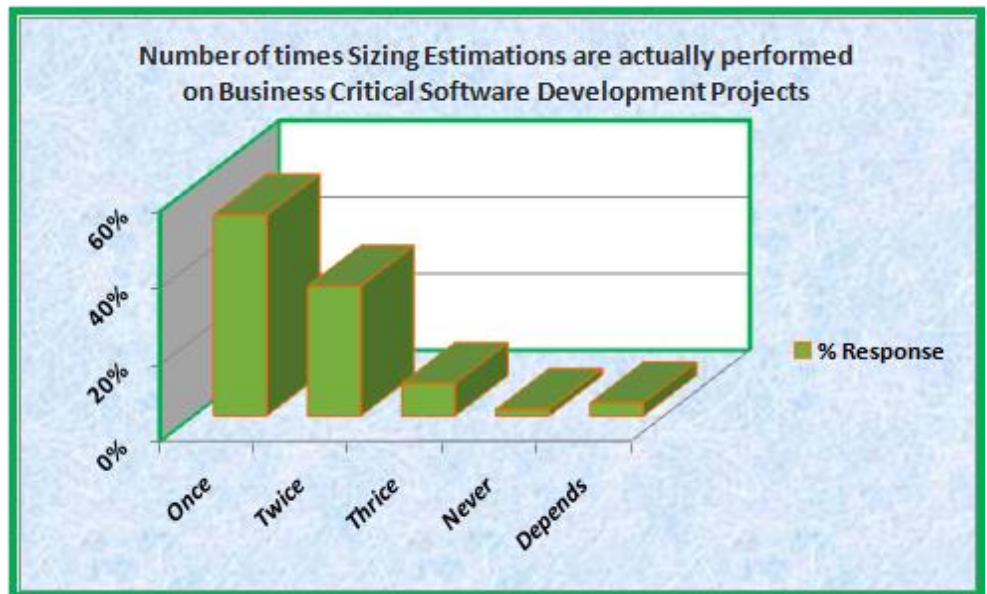

**Figure-4:** *Graphical representation of survey results*

A. From a look at the survey results, it is clearly evident that in the experience of real world project managers and practitioners, application of sizing estimation techniques are done only once or twice in more than 85% of the projects.

B. These results are aligned with Standish group's chaos report, which has found average schedule overrun of 80% in global IT-projects. This provides an indication that overruns may have linkage with estimation errors.

C. Learning from Phase-II of this study, which included detailed assessment and investigation of 15-large projects from a proper blend of multiple organizations (spanning matured organizations, start-ups and SMBs) is listed below. Projects for this phase of study were selected on the basis of high-degree of project's malfunctioning (schedule overruns and effort overruns) in consultation with respective organizational PMOs and/or PMs. Our findings from this research are as follows:

   a. In 100% of projects under the scope of this study (*11 out of 11*):
      i. Underestimation of size, cost & schedule was clearly found – when compared with actual software-size.
      ii. No evidence of using any estimation tool or estimation software was found.

   b. In 91% of projects under the scope of this study (*10 out of 11*), there is no 'risk management' being actually being practiced – in real sense.

   c. In 82% of projects under the scope of this study (*09 out of 11*):
      i. The sizing estimates available with project team were actually done using top-down approach. On all such projects, project teams were unable to defend their estimates – when challenged on accuracy of estimates, using bottom-up approach.
      ii. Early size estimates – were found to be converted into legally binding contractual commitments, without performing due diligence on qualifying and validating the proposed guesstimates.
      iii. Documented process for sizing-estimations, was neither available nor in pipeline.
      iv. There was no database of historical project delivery actuals and estimates were not being validated from organizational norms and productivity figures. Surprisingly, project teams were not even having any plan of maintaining historical database for archiving of project actual performance data, for future reference.

   d. In 73% of projects under the scope of this study (*08 out of 11*), sizing estimations were done only once at proposal stage at inception of the project and not repeated subsequently during life cycle of the project.

   e. In 18% of projects under the scope of this study (*02 out of 11*):
      i. Sizing estimations were claimed to have been validated against benchmarks, but it was found that organizational database meant for archiving project history is

      not being updated religiously with project's actuals. Consequently, the validation itself was not effective and meaningful.

    ii. Internal benchmarks of the organization were not being revised, in-line with historical projects performance data.

    iii. Estimates were provided by senior developers – who were not having any formal training on estimation tools and techniques. It was presumed that they would be good estimators.

  f. This research and study revealed that there is a widespread ignorance on inherent risks associated with sizing estimates of the project, irrespective of the size measure (be it SLOC or FSM) and associated risk mitigations needs.

  g. This research and study revealed that, in many projects, there is plenty of scope for improving the estimation validation and qualification process prior to converting the proposed estimates into contractual commitments.

  h. Our study also revealed that most dominant factor for cost estimation is the size estimation. Although there are other factors as well, but the cost estimation risk (i.e. the possibility that the estimated cost will be far different from the actual software cost) depends more on inaccurate size estimates than on any other cost-related parameter.

D. This study has resulted into a valuable contribution by way of proposing a simple, pragmatic and easy to deploy solution of <u>*sizing estimation stage-gates*</u>, to mitigate inherent risks associated with application of any of the existing sizing estimation techniques for software projects. This solution, articulated in Table-5, offers a prescriptive guidance for mitigating sizing estimation related risks for business critical software projects. It is recommended that contractual commitments (including final cost and final schedule) on business critical software projects should be deferred until ESG-03. Any commitments established prior to ESG-03, should be clearly confirmed as "tentative" or "draft" that is liable to change significantly.

E. Authors strongly believe and conclude that the proposed layered approach to application of sizing estimations to business critical software projects, using the stage-gates (Appendix-A: Table-5) will increase the size estimation accuracy for the projects and contribute significantly towards alleviation of the software crisis.

## 8. Conclusion

(i) An important finding of this study is identification and acknowledgement of widespread ill-practice of "*unqualified guesstimates, prepared by an individual, based solely on intuition, often gets rebranded as official project estimates and formal contractual commitments*" in many start-ups as well as SMBs.

(ii) Another important finding of this study is unearthing the need for organizations to have a defined process of estimations for business critical software projects – using some kind of quantitative

(ii) methods for estimations (preferably using bottom-up approach) and requirement of having a project database for archiving history information on projects, which should provide a valuable input to estimation process of the organization.

(iii) This study has acknowledgement of existence of risks in any software sizing estimate and offered a pragmatic solution for mitigating the risks caused by poor sizing estimates.

(iv) As with any research, this study also has limitations associated with methodology and data set. This study's limitations notwithstanding, it does provide evidence of direct effect of software sizing estimates on effort/cost and schedule estimates for business critical software projects.

(v) Sizing estimation errors are also likely to have indirect effects on cost and schedule overruns. Future studies, on this can measuring and model the correlations between size underestimations and effort/cost/schedule overruns.

## 9. Implications For Further Research

This research is a step forward towards further understanding of the impact of sizing estimation errors on prevailing software crisis for business critical software projects. Upon acknowledging the existence of risks, this research has focused on providing a groundwork for a new layered approach to application of software sizing estimation techniques on business critical software projects, in order to minimize the risks associated with sizing estimation errors. Planned extensions of work, in this regard could be:

- Deploying the proposed layered approach in real world projects and measure the benefits quantitatively.

- Taking-up industry case-studies on developing mathematical models for correlation between size underestimates and cost/schedule performance of software projects.

- Taking-up industry case study on automation of estimation process, using estimation tools and measuring the ROI-benefits.

### Acknowledgements

Authors would like to thank all the organizations, project sponsors, program managers, project managers and practitioners from Global IT-Industry who participated in the survey conducted during Phase-I of this study. Also, special thanks to senior project managers and team members of the projects selected for detailed study for their cooperation, support and sparing their time, providing their valuable inputs as and when required during the course of detailed study, including semi-structured interviews that were conducted and iterated to weed-out the results. Many thanks to all the peers, colleagues and organizations who supported this research by sharing their valuable time and feedback on experiences in application of sizing estimations techniques during life cycle of business critical software projects.

## Appendix-A

Table-5: Sizing Estimation Stage Gates for Business Critical Software Project Management

| Sr. No. | Estimation Stage-Gate | When should it be done during Project Life Cycle | Utility of Estimate for Bidders | Utility of Estimate for Consumers | How to Validate the Estimate | Who should do the estimation |
|---|---|---|---|---|---|---|
| 1. | ESG-01 | Early Size Estimation (*at proposal stage of the project*) | (i) An early-size-estimation is used to scope the project during bidding. (ii) It provides a general sense of how large the proposed system may be, with a view of high-level planning of project resources (including people and time) that may be needed to build the system and thus providing essential input to respond to requests for tenders/ Technical Bids/quotations. | To determine whether the bids are reasonable, cost analysts can evaluate the bidders' | Reviewer should check the sizing method and detailed worksheet for appropriateness and accuracy by examining the inputs and estimates prepared.  Compare them with similar projects in the project database of historical projects, delivered by the organization (if such a database exists). | Proposal Team (pre-sales team) |
| 2. | ESG-02 | Estimation during requirements elicitation phase of SDLC | (i) To manage user expectations about each requirement, in terms of what adding a requirement will mean to the overall project size, cost, schedule & risk. (ii) To weigh priorities among requirements, helping to avoid "*gold-plating*" some requirements (i.e. adding unnecessary or unnecessarily complex requirements) that can be achieved more simply and cheaply. | To assess financial and schedule implications of adding additional requirements, thereby taking informed decisions on requirements baseline. | (i) Acquire multiple independent estimation details, in the context of elaborated requirements and confront the gaps till the estimates converge. (ii) Use multi-point estimates, instead of single point estimates and compare with other projects actual sizing data from organizational | By Project Delivery Organization |

| Sr. No. | Estimation Stage-Gate | When should it be done during Project Life Cycle | Utility of Estimate for Bidders | Utility of Estimate for Consumers | How to Validate the Estimate | Who should do the estimation |
|---|---|---|---|---|---|---|
| | | | | | history data base (if available). | |
| 3. | ESG-03 | Estimation during solution design phase of SDLC | (i) To make informed decisions on technical design alternatives after evaluating multiple design options within the boundaries of projects triple constraints | To assess financial, schedule and operational implications of alternative design options, thereby taking active part in design decisions by representing users perspective. | (i) Acquire multiple independent estimation details, in the context of technical design options and confront the gaps till the estimates converge.<br>(ii) Use multi-point estimates, instead of single point estimates and compare with other projects actual sizing data from organizational history data base (if available). | By Project Delivery Organization |
| 4. | ESG-04 | Estimation during construction phase of SDLC | (i) During the course of development, measuring size of code completed and estimating the remaining size is essential for project managers to predict the likely size of the finished product, and to determine whether the project is on track or not.<br>(ii) Consequent upon approved change requests during the course of project, the requirements and design may get changed. Re-estimating the size after requirements and design modifications have been approved is essential for | (i) To get an update on projects health by way of reviewing the actual size and remaining size to be constructed.<br>(ii) To assess the impact of change requests on project's triple constraints. | (i) Acquire multiple independent estimation details, in the context of actual progress, sizing attributes of change requests(if any) and confront the gaps till the estimates converge.<br>(ii) Use multi-point estimates, instead of single point estimates and compare with other projects actual sizing data from | By Project Delivery Organization |

| Sr. No. | Estimation Stage-Gate | When should it be done during Project Life Cycle | Utility of Estimate for Bidders | Utility of Estimate for Consumers | How to Validate the Estimate | Who should do the estimation |
|---|---|---|---|---|---|---|
| | | | re-baselining project. (iii) Size tracking and re-estimation during development phase is helpful to evaluate productivity (*not only to determine whether the project team needs help but also to assist in decisions about reorganizing project resources*) | | organizational history data base (if available). | |
| 5. | ESG-05 | Estimation during implementation phase of SDLC | Size Estimations during implementation phase is a very handy tool for deploying, stabilizing handover of the system to support organization for warranty support, prior to releasing project teams. | To help stakeholders assess whether the system being implemented is far more or less ambitious than was envisioned earlier in the project. | (i) Acquire multiple independent estimation details (including at least one size estimate from support team), in the context of application deployed and being transitioned to support teams and confront the gaps till the estimates converge. (ii) Use multi-point estimates, instead of single point estimates and compare with other projects actual sizing data from organizational history data base (if available). | Jointly by Project Delivery Organization and Post-Project Support Organization |
| 6. | ESG-06 | Estimation during Project Closure Phase | Size Estimations during project closure phase is an essential item for capturing the lessons learnt and updating | To get an update on actual size of system delivered to | | By Project Delivery Organization |

| Sr. No. | Estimation Stage-Gate | When should it be done during Project Life Cycle | Utility of Estimate for Bidders | Utility of Estimate for Consumers | How to Validate the Estimate | Who should do the estimation |
|---|---|---|---|---|---|---|
| | | | project history databases and organizational process assets. | business – for future purposes. | | |